\documentclass[12pt]{iopart}

\usepackage{graphicx}

\begin{document}

\bibliographystyle{unsrt}

\title[Ground-state configuration space heterogeneity of random spin glass models]{Ground-state configuration space heterogeneity of random finite-connectivity spin glasses and random constraint satisfaction problems}

\author{Haijun Zhou$^{1,2}$ and Chuang Wang$^{1,3}$}

\address{$^1$Key Laboratory of Frontiers in Theoretical Physics,
Institute of Theoretical Physics, Chinese Academy of Sciences, 
Beijing 100190, China}

\address{$^2$Kavli Institute for Theoretical Physics China (KITPC) at the
Chinese Academy of Sciences, Beijing 100190, China}

\address{$^3$School of Physics, Northeast Normal University,
Changchun 130024, China}

\ead{zhouhj@itp.ac.cn, chuangphy@gmail.com}

\begin{abstract}
We demonstrate through two case studies, one on the $p$-spin interaction 
model and the other on the random $K$-satisfiability problem, that a
heterogeneity transition occurs to the ground-state configuration
space of a random finite-connectivity spin glass system
at certain critical value of the constraint density.
At the transition point, exponentially many configuration communities
emerge from the ground-state configuration space, making the
entropy density $s(q)$ of configuration-pairs a non-concave function of
configuration-pair overlap $q$. Each configuration community is a collection of relatively
similar configurations and it forms a stable thermodynamic
phase in the presence of a suitable external field.
We calculate $s(q)$ by the replica-symmetric
and the first-step replica-symmetry-broken cavity methods,
and show by simulations that the configuration space
heterogeneity leads to
dynamical heterogeneity of particle diffusion processes
because of
the entropic trapping effect of configuration communities. This work
clarifies the fine structure of the ground-state configuration space
of random spin glass models, it also sheds light on the 
glassy behavior of hard-sphere colloidal systems at relatively 
high particle volume fraction.
\end{abstract}




\maketitle


\section{Introduction}
\label{sec:introduction}

Spin glass models defined on finite-connectivity random graphs have two
control parameters, one is the temperature $T$ and the other is the
interaction (or constraint) density, defined as the number $M$ of interactions 
(constraints) versus the number $N$ of vertices, $\alpha \equiv M/N$.
Theoretical work of the last ten years
\cite{Mezard-Parisi-2001,Mezard-Montanari-2006,Krzakala-etal-PNAS-2007}
have established the understanding that, a random spin glass system
with many-body interactions will experience a clustering transition
if either the temperature $T$ is lowered below certain threshold value
or the constraint density $\alpha$ is increased beyond certain threshold
value.  At the clustering transition, the
configuration space of the system splits into an exponential
number of Gibbs pure states and ergodicity is broken.  This
transition is followed by another phase transition of
the configuration space, the condensation transition,
as $T$ is further lowered or $\alpha$ further increased.
At the condensation transition, a sub-exponential number of
large Gibbs states start to
dominate the configuration space and hence the equilibrium property of
the system \cite{Krzakala-etal-PNAS-2007}. These properties were first
observed in fully-connected mean-field spin glass models
\cite{Gardner-1985}.  Whether they are also valid in $D$-dimensional 
real-world spin glass systems ($D=3$ or $D=2$) is still a debated open issue.

Although the configuration space geometric properties of random
finite-connectivity spin glass models at or after the clustering
transition have been well characterized by statistical physics
methods, much less is known about the configuration space structure
just before the clustering transition. Why do an exponential number of
Gibbs states suddenly appear at the clustering transition?
Are they preceded by precursor structures in the ergodic phase of
the configuration space? If yes, when do the Gibbs-state precursors
start to form and how to describe their evolution? What are the
impacts of these precursor structures to the equilibrium dynamical
properties of the system? These questions
are important for a full understanding of the structural evolution of
the configuration space.

In this paper,
as a continuation of our recent efforts
\cite{Li-Ma-Zhou-2009,Zhou-Ma-2009,Zhou-2009-b,Zhou-2010},
we study the fine structure of the configuration space
of random finite-connectivity spin glass models in the vicinity of the
clustering transition.
We focus on the evolution of the
ground-state configuration space (corresponding to $T=0$)
using the constraint density $\alpha$ as the control parameter.
As demonstrated through two case studies, a
heterogeneity transition occurs to the ground-state configuration
space at certain critical value of $\alpha$.
At the transition point, exponentially many communities
of configurations 
emerge from the ground-state configuration space.
Each configuration community is a collection of relatively
similar configurations and it forms a stable thermodynamic
phase in the presence of a suitable external field; they are the
precursors for the Gibbs states at the clustering transition.
The entropy density $s(q)$ of configuration-pairs
as a function of configuration-pair overlap $q$
is calculated by the replica-symmetric
(RS) and the first-step replica-symmetry-broken (1RSB)
cavity methods. Extensive numerical simulations are performed to
confirm that, the entropic trapping effect of ground-state
configuration communities leads to strong dynamical heterogeneity
of diffusion processes within the configuration space.

Dynamical heterogeneity in the fully-connected $p$-spin interaction
spherical model (with continuous spins)
\cite{Crisanti-Sommers-1992} was quantitatively
investigated by Donati and co-workers \cite{Donati-etal-2002} within
the framework of the Franz-Parisi effective potential theory
\cite{Franz-Parisi-1995,Franz-Parisi-1997}. The results of the
present paper for finite-connectivity systems with discrete spins
are qualitatively similar to the results of \cite{Donati-etal-2002}.
Some aspects of the heterogeneity transition in
random finite-connectivity systems are also shared by real-world
$D$-dimensional spin glass and structural glass systems, such as
hard-sphere poly-disperse colloidal systems, where the particle
volume fraction plays the role of the constraint density $\alpha$
\cite{Donati-etal-2002}. In real-world glass systems and supercooled
liquid systems, dynamic heterogeneity occurs in real space.
The present work is also related to the work of Krzakala and
Zdeborova on the adiabatic evolution of a single Gibbs state of a
finite-connectivity spin glass system as a function of temperature
$T$ \cite{Krzakala-Zdeborova-2009b,Zdeborova-Krzakala-2010}. 
 
Two model systems are studied in the paper. Section \ref{sec:kxorsat}
concerns with the $p$-body spin glass system, which
is equivalent to the random $K$-XORSAT (exclusive-or-satisfiability)
problem of computer science. Replica-symmetric and 1RSB mean-field
calculations are carried out to obtain the entropy density of
ground-state configuration-pairs. Section \ref{sec:ksat} focuses on the heterogeneity
transition and the dynamical heterogeneity of the random
$K$-satisfiability ($K$-SAT) problem. Results obtained by
RS calculations and numerical simulations are reported
in this section. We make further discussions in section
\ref{sec:discuss}.

\section{The random $p$-body spin glass model}
\label{sec:kxorsat}

The random $p$-body spin glass model is defined by the
energy function
\begin{equation}
	E_{pspin}( \vec{\sigma} ) = - \sum\limits_{a=1}^{M}
	 J_a \prod\limits_{i \in \partial a}  \sigma_{i} \ ,
	\label{eq:p-body}
\end{equation}
where $\vec{\sigma} \equiv (\sigma_1, \sigma_2, \ldots, \sigma_N)$
denotes a spin configuration for the $N$ vertices $i\in [1, N]$, with
each spin variable $\sigma_i \in \{-1, +1\}$; 
the index $a\in [1, M]$ denotes one of the $M$
interactions of the system and $J_a$ is the quenched random coupling constant,
whose value is fixed to $J_a=+1$ or $J_a=-1$ with
equal probability; the set $\partial a$ includes all the
vertices that participate in the interaction $a$, 
its size is fixed to $p$, and each of
its $p$ different elements is randomly and
uniformly chosen from the whole set
of $N$ vertices. Similar to $\partial a$,
we denote by $\partial i$ the set of interactions
that involve vertex $i$.
While each interaction $a$ in the $p$-body spin glass model
affects the same number $p$ of vertices, the sets $\partial i$ may
have different sizes for different vertices $i$. Actually, when $N$ is
large enough, the probability
that a randomly chosen vertex participates in $k$ interactions is
governed by the Poisson distribution $P_P(k) = e^{-c} c^{k} / k!$,
with mean value $c = p M/N$.

In the $p$-body spin glass model \eref{eq:p-body},
each interaction contributes either a positive energy
$+1$ or a negative energy $-1$ to the total energy.
If we set $p=K$, this model is 
equivalent to the random
$K$-XORSAT problem with the energy function
\begin{equation}
     E_{xorsat}( \vec{\sigma} ) = \sum\limits_{a=1}^{M}
     \frac{1-J_a \prod_{i\in \partial a} \sigma_{i}}{2} \ .
	\label{eq:k-xorsat}
\end{equation}
For the random $K$-XORSAT problem, each interaction $a$ is also referred
to as a constraint, whose energy
 $(1-J_a \prod_{i\in \partial a} \sigma_i)/2$
 is either zero (constraint being satisfied) or unity (constraint
violated). The energy $E_{xorsat}(\vec{\sigma})$ counts the
total number of violated constraints by the spin configuration
$\vec{\sigma}$. The constraint density $\alpha$ of the system is
by $\alpha = M/N$.

The random $K$-XORSAT problem (or equivalently, the random $p$-body
spin glass model) has been well studied in the statistical physics
community. It serves
as an interesting system for understanding the low-temperature
equilibrium property of finite-connectivity spin glasses
\cite{Mezard-Parisi-2001,RicciTersenghi-Weigt-Zecchina-2001,Mezard-etal-2003,Cocco-etal-2003},
and for understanding the dynamical property of glassy systems
\cite{Montanari-Semerjian-2006,Montanari-Semerjian-2006b}.
This model is also closely related to error-correcting code systems
of information science, such as the Sourlas code \cite{Sourlas-1989}. 
The ground-state configuration space structure of the random $K$-XORSAT
problem has been investigated in great detail
\cite{Mezard-etal-2003,Cocco-etal-2003,Mora-Mezard-2006}
and was found to depend only on $K$ and the constraint density $\alpha$
in the limit of  $N\rightarrow \infty$.

For a given value of $K\geq 2$ there is a satisfiability threshold
$\alpha_s(K)$. When the constraint density $\alpha$ is below $\alpha_s(K)$,
the ground-state energy of model \eref{eq:k-xorsat}
is zero (the system is in the SAT phase), but it
becomes positive when $\alpha > \alpha_s(K)$ (the UNSAT phase).
A zero-energy spin configuration of model
\eref{eq:k-xorsat} is referred to as a solution, and all
the solutions form the solution space $\mathcal{S}$ of this system.
The solution space
of a large random $K$-XORSAT problem is non-empty only if its constraint
density is in the range of $\alpha \leq \alpha_s(K)$.
We have $\alpha_s(2)=0.5$,
$\alpha_s(3)=0.918$,  and
$\alpha(4)=0.977$ \cite{Mezard-etal-2003}.

Before the solution space of the random $K$-XORSAT ($K\geq 3$)
becomes empty at $\alpha > \alpha_s(K)$,
it experiences an ergodicity-breaking
(clustering) transition at the threshold value $\alpha= \alpha_d(K)$,
with $\alpha_d(3)=0.818$ and $\alpha_d(4)=0.772$ \cite{Mezard-etal-2003}.
For $\alpha < \alpha_d(K)$, the whole solution space
forms a single Gibbs state (the meaning of a Gibbs state is
explained geometrically in the following subsection). On the other hand,
for $\alpha > \alpha_d(K)$, the solution space is no longer ergodic but
is composed of exponentially many solution clusters (Gibbs states),
each of which containing exactly the same number of solutions.
For $\alpha_d(K) < \alpha < \alpha_s(K)$, solutions from different
solution clusters are separated by energy barriers that are proportional
to the vertex number $N$ \cite{Montanari-Semerjian-2006}.

\subsection{Non-concavity of the entropy function and solution space
heterogeneity}

\begin{figure}
 \begin{center}
        \includegraphics[width=0.7\textwidth]{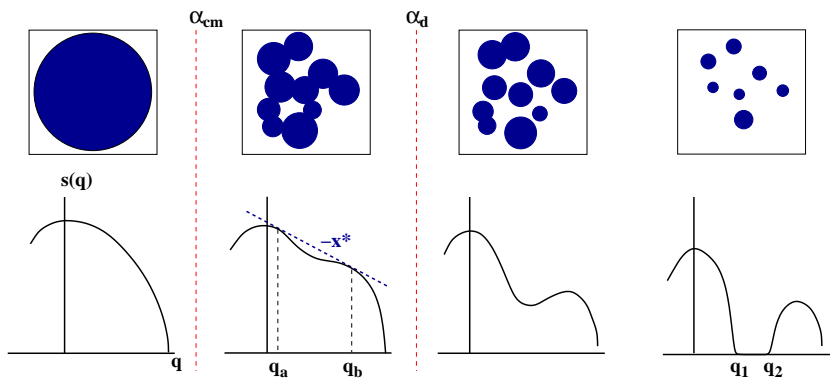}
    \end{center}
    \caption{\label{fig:community}
    Non-concavity of the entropy function $s(q)$ as defined by \eref{eq:entropy}
    and the large scale organization of the solution space of a random
    constraint satisfaction problem. $s(q)$ is a concave
    function of the solution-pair overlap $q$ when the constraint density
    $\alpha$ is low enough (left-most panel); at $\alpha = \alpha_{cm}$,
    $s(q)$ starts to be non-concave, marking the formation of
    many solution communities in the solution space (second panel from left);
    then as $\alpha$ increases to $\alpha= \alpha_d$, $s(q)$ becomes
    non-monotonic, the solution communities separate into different
    solution clusters and ergodicity of the solution space is broken;
    the solution clusters shrink in size as $\alpha$
    further increases, and eventually the number of solution-pairs with
    intermediate overlap values $q\in [q_1, q_2]$ will be zero 
    and $s(q)$ is only positive for $q> q_2$ and $q< q_1$.
    }
\end{figure}

The similarity of two solutions $\vec{\sigma}^1$,
$\vec{\sigma}^2$ in the non-empty solution space $\mathcal{S}$
of a random $K$-XORSAT problem can be measured by the overlap value
\begin{equation}
	q(\vec{\sigma}^1, \vec{\sigma}^2)
	= \frac{1}{N} \sum\limits_{i=1}^{N} \sigma_i^1 \sigma_i^2 \ .
	\label{eq:overlap}
\end{equation}
We denote by $\mathcal{N}(q)$ the total number of solution-pairs
in the solution space $\mathcal{S}$ with an overlap value $q$. This number
is exponential in $N$ in the SAT phase $\alpha \leq \alpha_s(K)$.
Therefore in the limit of large $N$,
an entropy density $s(q)$ is defined as
\begin{equation}
	s(q) = \frac{1}{N} \ln \mathcal{N}(q) \ .
	\label{eq:entropy}
\end{equation}
The entropy density function $s(q)$ contains rich information about the
structure of the solution space $\mathcal{S}$
(see figure~\ref{fig:community}). The shape of $s(q)$ has several
qualitative changes as the constraint density $\alpha$ increases.
The first qualitative change occurs at $\alpha=\alpha_{cm}$, where
$s(q)$ becomes non-concave. This concavity change
corresponds to the formation of (exponentially) many solution communities
in the solution space (the large-scale homogeneity of the solution space
is then broken) \cite{Zhou-Ma-2009,Zhou-2009-b}.
We refer to $\alpha_{cm}$ as the heterogeneity transition point.

We introduce the following partition function $Z(x)$
for the solution space $\mathcal{S}$ \cite{Zhou-2009-b}
\begin{equation}
	\label{eq:Partition_function}
	Z(x) = \sum\limits_{\vec{\sigma}^1 \in \mathcal{S}}
	\sum\limits_{\vec{\sigma}^2 \in \mathcal{S}}
	\exp\Bigl( x \sum\limits_{i=1}^{N} \sigma_i^1 \sigma_i^2 \Bigr)
	= \sum\limits_{q} \exp\Bigl[ N \bigl(
	s(q) + x q \bigr) \Bigr] \ ,
\end{equation}
where $x$ is a coupling field between the solutions. When $x$ is positive,
solution-pairs with larger overlap values have larger weights in the
partition sum \eref{eq:Partition_function}.
Under the field $x$, the mean solution-pair overlap value
is $\overline{q}(x)= (1/N){\rm d} \ln Z(x)/ {\rm d} x$. For $N\rightarrow
\infty$, we see from \eref{eq:Partition_function} that,
$\overline{q}(x)=\arg\,\max_{q} \bigl( s(q)+ x q\bigr)$.
 At $x=0$, $\overline{q}(0)$ is equal to $q_0$,
the most probable solution-pair overlap value of the solution space.

If the entropy density $s(q)$ is concave in $q\in [q_0, 1]$, then
$\overline{q}(x)$ increases continuously with $x$ for $x \geq 0$.
However if $s(q)$ is non-concave, because there exists a line of slope
$-x^*$ which touches $s(q)$ at two different points $q=q_a$ and
$q=q_b> q_a$ (\fref{fig:community}), then
the value of $\overline{q}(x)$ jumps
from $q_a$ to $q_b$ at $x=x^*$ \cite{Zhou-Ma-2009,Zhou-2009-b}.  
This discontinuity of $\overline{q}(x)$ reveals the existence of
a field-induced first-order phase transition at $x=x^*$. The solution-pairs
exhibit two different levels of similarity.  For $x> x^*$, the
partition function $Z(x)$ are contributed overwhelmingly by
(intra-community) solution-pairs with overlap values $\geq q_b$ 
(overlap-favored phase),
while for $x<x^*$, $Z(x)$ is dominated by
(inter-community) solution-pairs with overlap values $\leq q_a$
(entropy-favored phase). The difference $q_b-q_a$ increases from
zero as the constraint density $\alpha$ exceeds the critical value
$\alpha_{cm}$. At $\alpha= \alpha_{cm}$, the solution space is in
a critical state, where the boundaries between different solution
communities are elusive, and the field-induced phase transition is
second-order.

The Hamming distance between two solutions $\vec{\sigma}^1$ and
$\vec{\sigma}^2$ is defined as $D(\vec{\sigma}^1, \vec{\sigma}^2)
\equiv \sum_{i=1}^N (1- \sigma_i^1 \sigma_i^2)/2$, which 
is related to the solution-pair overlap by $D(\vec{\sigma}^1,
\vec{\sigma}^2) = (N/2) (1-q(\vec{\sigma}^1, \vec{\sigma}^2) )$.
The solution space $\mathcal{S}$ can be represented by a graph of
nodes and edges. Each node of this solution graph
denotes a solution, and an edge is linked between two nodes of the
graph if the corresponding two solutions has a Hamming distance
not exceeding a specified value $D_0$.
For the random $K$-XORSAT problem at
$\alpha< \alpha_d(K)$, there exists a minimum value of $D_0 \ll N$
such that all the nodes of the solution graph are in a single
connected component \cite{Montanari-Semerjian-2006}.
We take this minimum value as our
edge linking criterion. (For $\alpha> \alpha_d(K)$, if $D_0$ is
not of the same order as $N$, the solution graph will be a collection
of exponentially many disjointed components.)
In the ergodic phase of $\alpha < \alpha_d(K)$, we may introduce
two particles to the solution graph. Initially these two particles
are residing on the same node, say $\vec{\sigma}^0$, of the graph.
In case the particles are uncoupled, then each particle
performs a random diffusion in the solution graph independent of
the other: Suppose at time $t$ the particle is at node $\vec{\sigma}$,
then at the next time step it will, with probability
$k_{\vec{\sigma}}/k_{max}$, make a move to a randomly chosen
nearest-neighbor of this node, where $k_{\vec{\sigma}}$
is the number of attached edges (the degree) of node $\vec{\sigma}$,
and $k_{max}$ is the maximal node degree in the solution graph.
In the case the particles are coupled by a field $x$, however,
the particle diffusions are mutually influenced, and the visited
node-pairs (say $\vec{\sigma}^1$ and $\vec{\sigma}^2$) are no longer
uniformly distributed but are favored to more similar pairs by a
factor of $e^{x N q(\vec{\sigma}^1, \vec{\sigma}^2)}$. In the
thermodynamic
limit $N\rightarrow \infty$, when the
coupling field $x$ is larger than $x^*$, then even at time approaching
infinity, the two particles will still be diffusing in the
neighborhood of each other and in the neighborhood of the initial
node $\vec{\sigma}^0$. Such a strong memory effect at
field value $x >x^*$ is a dynamical
manifest of the existence of communities in the solution space.

\subsection{Annealed approximation for $s(q)$}

If one knows a solution $\vec{\sigma}^1=(\sigma_1^1, \sigma_2^1,
\ldots, \sigma_N^1)$ for the $K$-XORSAT system \eref{eq:k-xorsat}, it
is convenient to perform a gauge transform ${\sigma}_i \leftarrow \sigma_i 
\sigma_i^1$ to the spin value of each vertex $i$. Under this transform,
\eref{eq:k-xorsat} is simplified to
\begin{equation}
	E(\vec{\sigma}) = \sum\limits_{a=1}^{M} \frac{1- \prod_{i\in \partial a}
	\sigma_i }{2} \ .
	\label{eq:k-xorsat-t}
\end{equation}
In this transformed system, all the coupling constants
are positive ($J_a=+1$), and the overlap of the
transformed solution $\vec{\sigma}$ with the reference solution
$\vec{\sigma}^1$ is $q(\vec{\sigma})=(1/N) \sum_i \sigma_i$.
\Eref{eq:k-xorsat-t} is independent of the reference solution
$\vec{\sigma}^1$. This is a well-known property of the $K$-XORSAT problem,
namely its solution space $\mathcal{S}$ has the same local and global
structure when viewed from any of its solutions. Because of this nice
property, instead of calculating the solution-pair number $\mathcal{N}(q)$,
we calculate the number $\mathcal{N}_1(q)$ of solutions which have an overlap
value $q$ with a reference solution $\vec{\sigma}^1$.
We denote the corresponding entropy density also as $s(q)$, i.e.,
$s(q) = (1/N)\ln \mathcal{N}_1(q)$.
(With this slight abuse of notation,
the solution-pair entropy density as defined
by \eref{eq:entropy} is $s(q)+s_0(\alpha)$,
where $s_0(\alpha)$ is the entropy density of the whole solution space at 
constraint density $\alpha$.) 

\begin{figure}
 \begin{center}
	\vskip 0.82cm
        \includegraphics[width=0.6\textwidth]{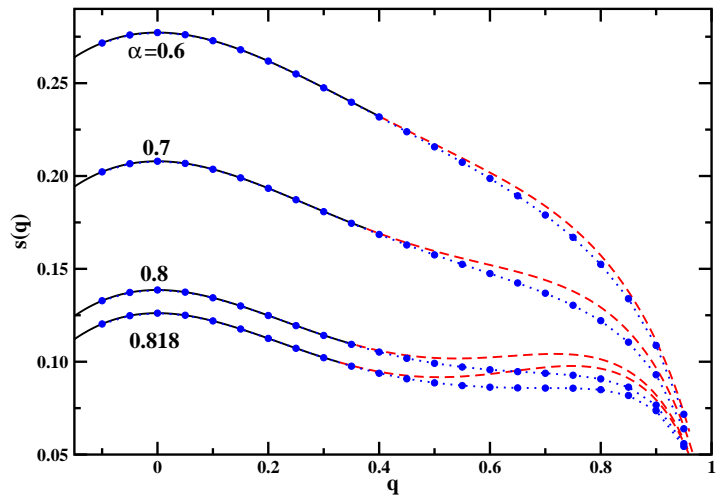}
    \end{center}
    \caption{\label{fig:anneal_xorsat}
     The entropy density $s(q)$ of solutions at an overlap level $q$ 
     to a reference solution for the random $K$-XORSAT problem with $K=3$
     and $\alpha=0.6$, $0.7$, $0.8$, $0.818$ (from top to bottom).
     Black solid lines and red dashed lines are results obtained by the
     annealed approximation \eref{eq:anneal-xorsat}, and blue dots are
     results of the replica-symmetric cavity method (dotted lines just for
     guiding the eye). When \eref{eq:consistency-xorsat} is valid,
     expression \eref{eq:anneal-xorsat} gives the exact mean value of $s(q)$,
      but if it is violated, \eref{eq:anneal-xorsat}
     is only an upper-bound for the mean value of $s(q)$.
     }
\end{figure}

The average value of $\mathcal{N}_1(q)$ over the ensemble of
random $K$-XORSAT problems with fixed vertex number $N$
and constraint density $\alpha$ is
\begin{equation}
 \overline{\mathcal{N}_1(q)}
= \left(
\begin{array}{c}
 N \\
(q+1) N/2
\end{array}
\right)
\left( \frac{1+q^K}{2}
\right)^{\alpha N}
\end{equation}
For $N\rightarrow \infty$ we obtain the following annealed approximation for
$s(q)$ as
\begin{equation}
	\label{eq:anneal-xorsat}
	s_{ann}(q)\equiv
\frac{\ln \overline{\mathcal{N}_1(q)}}{N}
 =-\frac{1+q}{2} \ln \frac{1+q}{2}
	-\frac{1-q}{2} \ln \frac{1-q}{2}
	+\alpha \ln \frac{1+q^K}{2} \ .
\end{equation}

For $K=3$, the function $s_{ann}(q)$ is concave in $q$ only
for $\alpha< 0.577$. For $0.577 < \alpha < 0.778$, $s_{ann}(q)$ is
non-concave in $q$ but is still monotonic in the range $q\in [0,1]$;
the monotonicity of $s_{ann}(q)$ in $q\in [0,1]$ is lost when $\alpha$
exceeds $0.778$ (\fref{fig:anneal_xorsat}). The same qualitative
results are obtained for the random $K$-XORSAT problem with $K \geq 4$.

The annealed approximation  $s_{ann}(q)$
is an upper bound of the entropy density $s(q)$
for a typical random $K$-XORSAT problem.
These two quantities are identical only when $\mathcal{N}_1(q)$ has the
self-averaging property, i.e., the value distribution of $\mathcal{N}_1(q)$
among the ensemble of random $K$-XORSAT problems of constraint density
$\alpha$ approaches a delta function in the limit of $N\rightarrow \infty$.
To check whether this self-averaging property
is valid, we need to calculate the mean value of $\mathcal{N}_1^2(q)$.
After some combinatorial analysis, we obtain
\begin{equation}
	\overline{\mathcal{N}_1^2(N)}
        =\left( \begin{array}{c}	N\\ (1+q)N/2 \end{array} \right)
        \sum\limits_{n} 
        \left( \begin{array}{c} (1+q)N/2 \\ n \end{array} \right)
        \left( \begin{array}{c} (1-q)N/2 \\ n \end{array} \right)
        Q_n^{\alpha N} \ ,
\end{equation}
where
$Q_n =[1+2 q^K + (1-4 n /N)^{K} ]/4$.
In the $N\rightarrow \infty$ limit, we therefore have
\begin{equation}
\frac{1}{N} \ln \overline{\mathcal{N}_1^2(q)}
=\max\limits_{0 \leq \rho \leq \min\left(\frac{1-q}{2}, \frac{1+q}{2}\right)}
s_2(q, \rho) \ ,
\end{equation}
with $s_2(q, \rho)$ being expressed as
\begin{eqnarray}
 s_2(q, \rho) = & -\frac{1+q}{2} \ln \frac{1+q}{2}
	-\frac{1-q}{2}\ln\frac{1-q}{2}-\rho \ln\frac{4 \rho^2}{1-q^2} 
\nonumber \\
&   -\left(\frac{1-q}{2}-\rho\right)\ln\left(1-\frac{2 \rho}{1-q}\right)
 -\left(\frac{1+q}{2}-\rho \right)\ln\left(1-\frac{2 \rho}{1+q} \right)
\nonumber \\
& +\alpha \ln\frac{1+2 q^K+(1-4 \rho)^K}{4} \ .
\end{eqnarray}

Self-averaging of $\mathcal{N}_1(q)$ requires that
\begin{equation}
\Delta \equiv \max\limits_{0 \leq \rho \leq \min\left(\frac{1-q}{2}, \frac{1+q}{2}\right)}
s_2(q, \rho) - s_{ann}^2(q) = 0  \ .
\label{eq:consistency-xorsat}
\end{equation}
Numerical calculations reveal that \eref{eq:consistency-xorsat} is
satisfied only for values of $q$ very
close to $0$ or very close to $1$. In \fref{fig:anneal_xorsat}, $s_{ann}(q)$ is
plotted as a black solid line if $\Delta <10^{-3}$ and as
a red dashed line if $\Delta \geq 10^{-3}$.

As $s_{ann}(q)$ is only an upper bound to $s(q)$, the fact that $s_{ann}(q)$
becomes non-concave at $\alpha < \alpha_d(K)$ can not be used as a proof
that $s(q)$ also becomes non-concave at $\alpha< \alpha_d(K)$.
We proceed to calculate
$s(q)$ by the cavity method of statistical physics.

\subsection{Replica-symmetric mean-field analysis}
\label{subsec:kxorsat-rs}

For the gauge-transformed $K$-XORSAT system \eref{eq:k-xorsat-t}, if
vertex $i$ is involved in constraint $a$, we define a cavity
probability $p_{i\rightarrow a}^{+}$ as
the probability that $\sigma_i$ takes the
value $\sigma_i=+1$ when the constraint $a$ is
absent. For each constraint $a$ we exploit the Bethe-Peierls approximation
\cite{Mezard-Parisi-2001,Kschischang-etal-2001,Mezard-Montanari-2009}
and assume that, the spin states of the vertices $j \in \partial a$ are
mutually independent in the absence of $a$.
Under this approximation, we obtain that
if the vertex $i\in \partial a$ takes the spin value $\sigma_i$,
the probability that constraint $a$ being satisfied is equal to
$[1+ \sigma_i \prod_{j\in \partial a \backslash i}
(2 p_{j\rightarrow a}^+ -1)]/2$, where
$\partial a\backslash i$ means the subset of $\partial a$ that is missing
element $i$. 
Then the following belief-propagation equation can be
written down for each vertex-constraint association
$(i, a)$:
\begin{equation}
\fl p_{i\rightarrow a}^{+}=
\hat{p}(\{ p_{j\rightarrow b}^+ \}) = 
 \frac{ e^{x} \prod\limits_{b\in \partial i \backslash a}
\left[ \frac{1+ \prod\limits_{j\in \partial b\backslash i}
 (2 p_{j\rightarrow b}^+-1)}{2} \right]}{
 e^{x} \prod\limits_{b\in \partial i \backslash a} \left[
\frac{1+ \prod\limits_{j\in \partial b\backslash i}
 (2 p_{j\rightarrow b}^+-1)}{2} \right]
+  e^{-x} \prod\limits_{b\in \partial i\backslash a} \left[
\frac{1- \prod\limits_{j\in \partial b\backslash i}
 (2 p_{j\rightarrow b}^+-1)}{2} \right]
} \ .
	\label{eq:bp-xorsat}
\end{equation}
For a given energy function \eref{eq:k-xorsat-t}, there are
$M\times K$ iteration equations, which form the replica-symmetric
cavity theory
\cite{Mezard-Parisi-2001,Kschischang-etal-2001,Mezard-Montanari-2009}.

\begin{figure}
 \begin{center}
	\vskip 0.82cm
        \includegraphics[width=0.6\textwidth]{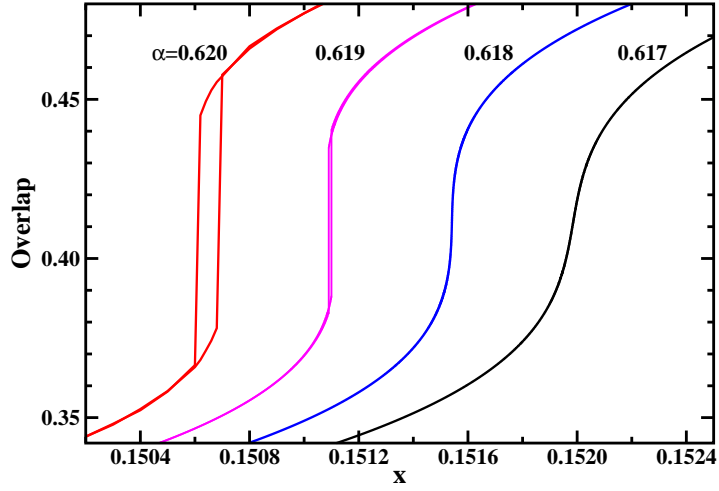}
    \end{center}
    \caption{\label{fig:q-xorsat-rs}
     The mean overlap $\overline{q}(x)$ for the random $3$-XORSAT problem at
     $\alpha = 0.617$, $0.618$, $0.619$, $0.620$ (from right to left).
     For $\alpha<\alpha_{cm}(K=3)=0.618154$, $\overline{q}(x)$ is a continuous
     function of $x$. At $\alpha=\alpha_{cm}(3)$,
     the slope of $\overline{q}(x)$ diverges at $x=0.151473$.
     For $\alpha> \alpha_{cm}(3)$, $\overline{q}(x)$
     is discontinuous and has a hysteresis loop.
     }
\end{figure}

After this set of belief-propagation iteration equations has
reached a fixed point, the mean overlap with the reference solution
is calculated as
\begin{equation}
	\overline{q}(x)= \frac{1}{N}
	\sum\limits_{i=1}^{N}
	\frac{ e^{x} \prod\limits_{a\in \partial i} \left[
\frac{1+ \prod\limits_{j\in \partial a\backslash i}
 (2 p_{j\rightarrow a}^+-1)}{2} \right]
-e^{-x} \prod\limits_{a\in \partial i} \left[
\frac{1 - \prod\limits_{j\in \partial a\backslash i}
 (2 p_{j\rightarrow a}^+-1)}{2} \right] }{
 e^{x} \prod\limits_{a\in \partial i} \left[
\frac{1+ \prod\limits_{j\in \partial a\backslash i}
 (2 p_{j\rightarrow a}^+-1)}{2} \right]
+  e^{-x} \prod\limits_{a\in \partial i} \left[
\frac{1- \prod\limits_{j\in \partial a\backslash i}
 (2 p_{j\rightarrow a}^+-1)}{2} \right]
} \ .
\label{eq:q-xorsat-rs}
\end{equation}
And the entropy density as a function of $x$ is
\begin{eqnarray}
\fl s(x) = \frac{1}{N}\sum\limits_{i=1}^{N} \ln\left(
e^{x} \prod\limits_{a\in \partial i} \left[
\frac{1+\prod\limits_{j\in \partial a\backslash i}
  (2 p_{j\rightarrow a}^+ -1)}{2} \right]
+ e^{-x} \prod\limits_{a\in \partial i} \left[
\frac{1-\prod\limits_{j\in \partial a\backslash i} (2 p_{j\rightarrow a}^+ -1)
}{2} \right] \right) \nonumber \\
 - \frac{1}{M}\sum\limits_{a=1}^{M} (K-1) \ln \left( \frac{
1+ \prod\limits_{j \in \partial a} (2 p_{j\rightarrow a}^+ -1) }{2} \right)
- x \overline{q}(x) \ .
\label{eq:s-xorsat-rs}
\end{eqnarray}

The mean value of $s(x)$ as averaged over the ensemble of random $K$-XORSAT
problems (at fixed value of $\alpha$)
 can be calculated using the population dynamics
technique \cite{Mezard-Parisi-2001,Mora-Mezard-2006}.
By eliminating $x$ from $s(x)$ and $\overline{q}(x)$ we obtain the entropy
density function $s(q)$. The numerical results are
shown in \fref{fig:anneal_xorsat} (blue dots) for $K=3$ at different
values of $\alpha$. The mean overlap function $\overline{q}(x)$ is shown in
\fref{fig:q-xorsat-rs} for $K=3$ and $\alpha\approx 0.62$. Similar results
are obtained for the cases of $K \geq 4$.

For the random $3$-XORSAT problem, $\overline{q}(x)$ is a continuous and
smooth function of field $x$ when $\alpha< \alpha_{cm}(3)\simeq
0.6182$. As $\alpha$ approaches $\alpha_{cm}$ from below, however,
the maximal slope of $\overline{q}(x)$ is proportional to
$(\alpha_{cm}(3)-\alpha)^{-1}$ and diverges at $\alpha_{cm}(3)$. This
divergence is a consequence of the fact that the entropy density $s(q)$
starts to be non-concave at $\alpha= \alpha_{cm}(3)$.
For $\alpha> \alpha_{cm}(3)$,
$\overline{q}(x)$ as calculated by the RS cavity theory
shows discontinuity and hysteresis  behavior when $x$ is close to certain
threshold value $x^*$, indicating the existence of
two distinct phases of the solution space as viewed from the reference
solution ($\vec{\sigma}^1$). One of the phases contains solution
$\vec{\sigma}^1$ and the other similar solutions, whose overlap with
$\vec{\sigma}^1$ is larger than certain characteristic value 
 $\approx (q_a+q_b)/2$, see \fref{fig:community}. We regard this phase as
the solution community of solution $\vec{\sigma}^1$. If we choose another
solution outside the solution community of $\vec{\sigma}^1$, we will find
that this new reference solution is also associated with a
different solution community. 

For $\alpha_{cm}(K) < \alpha < \alpha_{d}(K)$, the solution
space of the random $K$-XORSAT problem is therefore formed by exponentially
many solution communities. As the solution space is very heterogeneous at this
range of $\alpha$, the replica-symmetric mean-field
theory probably is not
sufficient to describe its statistical property. We now proceed to
study the solution space heterogeneity using the 1RSB cavity theory.

\subsection{First-step replica-symmetry-broken mean-field analysis}
\label{sec:1rsb-xorsat}

For the gauge-transformed model
\eref{eq:k-xorsat-t} under the coupling field $x$,
to apply the 1RSB mean-field theory, 
the solution space is first divided into an exponential
number of Gibbs states $\gamma$ \cite{Mezard-Parisi-2001,Mora-Mezard-2006}.
Each Gibbs state $\gamma$ represents a subspace
$\mathcal{S}_\gamma \subset \mathcal{S}$, 
and its  partition function is defined as
\begin{equation}
Z_\gamma (x) = \sum\limits_{\vec{\sigma} \in \mathcal{S}_\gamma}
\exp\left( x \sum_{i=1}^{N} \sigma_i \right) \ .
\end{equation}
We can then define a $`$free energy' density $f_\gamma(x)$ as
$f_\gamma = (1/N) \ln Z_\gamma(x)$. This free energy density is the sum of
two parts, $f_\gamma= s_\gamma+ x q_\gamma$, where
$s_\gamma$ is the entropy density of Gibbs state $\gamma$ and $q_\gamma$ is
the mean overlap level of solutions in $\mathcal{S}_\gamma$ to the reference
solution $\vec{\sigma}^1$.

\begin{figure}
 \begin{center}
        \includegraphics[width=0.6\textwidth]{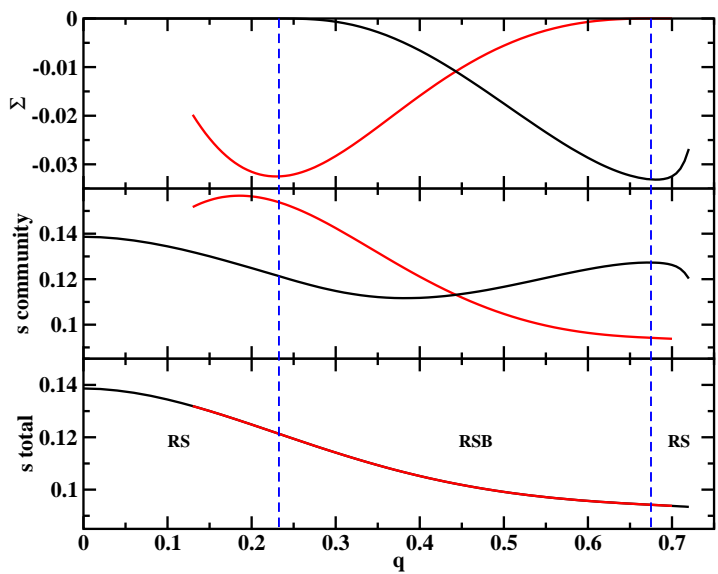}
    \end{center}
    \caption{\label{fig:sigma-xorsat-1}
     Results of the $m=1$ 1RSB mean-field theory for the random
     $3$-XORSAT problem with $\alpha=0.80$. The complexity,
     the entropy density of solution communities, and the total entropy
     density are shown in the upper, middle, and 
     lower panel, respectively,	as a function of the overlap $q$.
     Two sets of 1RSB mean-field results are
     obtained in the population dynamics when different types of
     initial values are used for the population.
     At each value of $q$, the result with the larger
     value of complexity and the smaller value of community entropy
     density should be considered. The two vertical dashed lines
     mark the boundary between the replica-symmetric (RS)  and the
     replica-symmetry-broken (RSB) region.
     }
\end{figure}

The total 1RSB partition function of the system is
\begin{equation}
\label{eq:z-1rsb}
Z_{1RSB} \equiv \sum\limits_{\gamma} (Z_\gamma)^{m} = 
\sum\limits_{\gamma} e^{N m f_\gamma} = \int {\rm d} f
\exp\left( N [ \Sigma(f) + m f ] \right)  \ .
\end{equation}
In the above equation, $m$ is the Parisi parameter, and
$\Sigma(f)$ is the complexity, which measures the $`$entropy density' of
Gibbs states at the free energy density level $f$
\cite{Mezard-Parisi-2001,Mora-Mezard-2006}.
If we set $m=1$ in \eref{eq:z-1rsb}, each Gibbs state contributes
a term $e^{N  f_\gamma}$ to $Z_{1RSB}$, and then $Z_{1RSB}$ is the total
partition function of the system, provided that the complexity $\Sigma(f)$
calculated at $m=1$ is non-negative. The existence of a nontrivial
1RSB solution at $m=1$ is also a signature of the instability
of the RS theory of the previous
subsection \cite{Mezard-Montanari-2006,Krzakala-etal-PNAS-2007}.
We therefore set $m=1$ in our calculations. The details of the 1RSB
cavity mean field theory are presented in \ref{sec:appendix-a},
and here we discuss some of the numerical results obtained by population
dynamics on the random $3$-XORSAT problem.

\begin{figure}
 \begin{center}
        \includegraphics[width=0.6\textwidth]{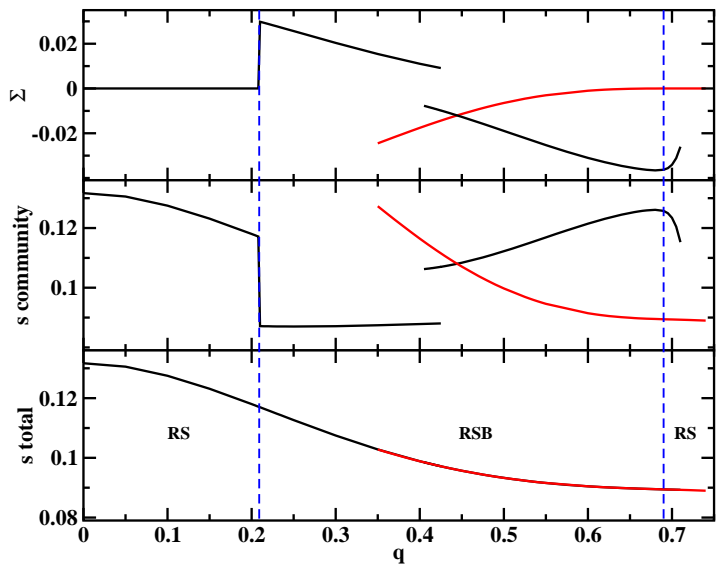}
    \end{center}
    \caption{\label{fig:sigma-xorsat-2}
     Same as \fref{fig:sigma-xorsat-1}, but for $\alpha=0.81$.
     The complexity $\Sigma(q)$ is positive for $0.21 \leq q \leq 0.425$.
     }
\end{figure}

At $\alpha < \alpha_{cm}(3)$, no nontrivial
1RSB mean field results are obtained. The fixed point of the 1RSB
population dynamics reduces to that of the RS theory
for all the overlap values $0 \leq q\leq 1$.
As $\alpha$ exceeds $\alpha_{cm}(3)$, the replica-symmetric mean field
theory becomes unstable for intermediate values of overlap $q$, and
nontrivial fixed points of the 1RSB mean-field population dynamics
are observed. For example, at $\alpha=0.80$ we find that the
complexity $\Sigma(q)$ is exactly zero for overlap value $q \leq 0.233$ and
$q \geq 0.675$, and $\Sigma(q)$ is negative for $0.233 < q < 0.675$
(see \fref{fig:sigma-xorsat-1}). The fact that $\Sigma(q)=0$ for
$q\geq 0.675$ suggests that, the solutions with overlap levels
$q\geq 0.675$ to $\vec{\sigma}^1$ are in
a single solution community. In the corresponding solution subgraph of
this solution community, any two solutions (nodes) with the
same Hamming distance $D$ to $\vec{\sigma}^1$ are connected by at least
one path that involves only other solutions with the same Hamming
distance $D$ to $\vec{\sigma}^1$ (i.e., the subspace of solutions having
the same overlap $q$ ($\geq 0.675$) with $\vec{\sigma}^1$ is ergodic within
itself). 

The second message we get, from the fact that $\Sigma(q)<0$ for
$0.233 < q < 0.675$, is that the solution subgraph formed by
all the solutions at the same  overlap level $q$ [$\in (0.233, 0.675)$] 
to $\vec{\sigma}^1$ is not ergodic within itself but is divided into
greatly many disjointed connected sub-components, with a
sub-exponential number of dominating ones. In this overlap range, the
community entropy density and the total entropy density as obtained
by the $m=1$ 1RSB mean-field theory can only be regarded as upper
bounds for the true values. One needs to work with $m<1$ to obtain
better estimates for the community entropy density and the
total entropy density.

The third message we get from $\Sigma(q)=0$ for $q\leq 0.233$ is that,
the solutions with  overlap levels to $\vec{\sigma}^1$ less than $0.233$
form an ergodic subspace within itself. More precisely, all the
solutions at each overlap level $q$ ($\leq 0.233$) to
$\vec{\sigma}^1$ are ergodic within themselves.
The solutions in such a subspace of fixed overlap $q$
come from different solution
communities, but they are connected with each other in the solution
graph even when only the edges inside the subgraph are remained.

\begin{figure}
 \begin{center}
        \includegraphics[width=0.6\textwidth]{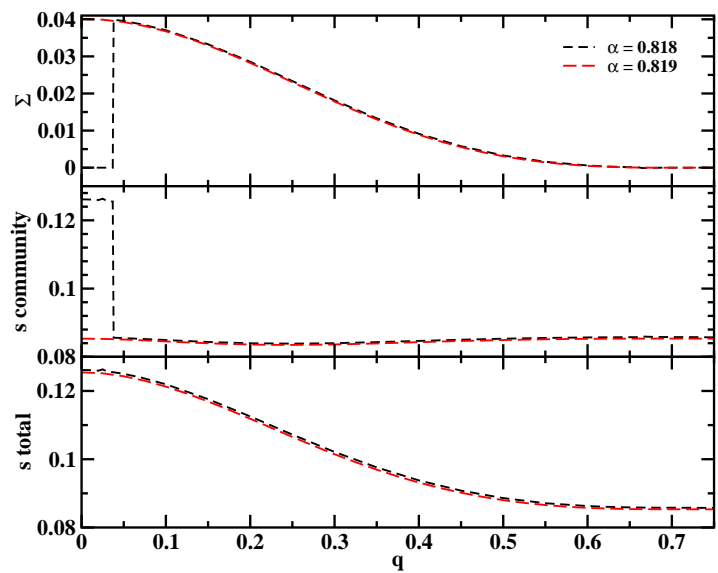}
    \end{center}
    \caption{\label{fig:sigma-xorsat-3}
     Same as \fref{fig:sigma-xorsat-1}, but for $\alpha=0.818$ and
     $\alpha=0.819$. At $\alpha=0.818$, the complexity $\Sigma(q)$
     is positive for $0.038 \leq q \leq 0.66$ and is negative
     ($\approx -10^{-5}$) for $0.66 < q \leq 0.705$. At
     $\alpha=0.819> \alpha_d(3)$, the complexity $\Sigma(q)$ is
     positive even at $q=0$.     
     }
\end{figure}

The theoretical results as shown in \fref{fig:sigma-xorsat-2}
and \fref{fig:sigma-xorsat-3} for the
random $3$-XORSAT problem at $\alpha=0.81$ and $\alpha=0.818$,
respectively, are similar with the results of
\fref{fig:sigma-xorsat-1}.  At $\alpha=0.81$,
the complexity $\Sigma(q)$
calculated at $m=1$ is positive for $0.21 \leq q \leq 0.425$,
indicating an exponential number of
solution communities are equally contributing to the solution
subspace at each overlap level $q \in [0.21, 0.425]$.
For an overlap level $q\in (0.425,  0.69$], however,
the solution subspace is again
dominated by a sub-exponential number of large solution
communities. The situation at $\alpha=0.818$ is similar,
but the range of $q$ values for which $\Sigma(q)>0$ is further
enlarged.

When the constraint density $\alpha$ exceeds the clustering
transition point $\alpha_d(3)$, the complexity $\Sigma(q)$ at $m=1$
becomes positive at $q=0$ (see the exemplar case of
$\alpha=0.819$ in \fref{fig:sigma-xorsat-3}). This
result means that
the largest subspace of solutions (with overlap level
$q=0$ to $\vec{\sigma}^1$) becomes non-ergodic at $\alpha>
\alpha_d(3)$, as expected.

\subsection{Summary for $K$-XORSAT}

The calculations of this section demonstrated
that the solution space of the random $K$-XORSAT problem
experiences a heterogeneity transition as the constraint density $\alpha$
approaches $\alpha_{cm}(K)$, where an exponential number of solution
communities start to form in the solution space. These solution communities
separate into different solution clusters at a larger constraint
density value $\alpha_d(K)$, where an ergodicity-breaking transition occurs.
For $\alpha_{cm}(K) < \alpha < \alpha_{d}(K)$, although the
solution space as a whole is ergodic, the subspaces of solutions with
intermediate overlap levels to a reference solution are non-ergodic
within themselves.

Our theoretical results, combined with the results of
\cite{Mora-Mezard-2006} for $\alpha > \alpha_d(K)$, give a
complete picture on the structural evolution of the solution space of
the random $K$-XORSAT problem.

\section{The random $K$-satisfiability problem}
\label{sec:ksat}

The random $K$-SAT problem is a famous model system
for the study of typical-case computational complexity of
NP-complete combinatorial satisfaction problems
\cite{Kirkpatrick-Selman-1994}.
Its energy function $E(\vec{\sigma})$, like the random
$K$-XORSAT problem, is defined as a sum
over $M= \alpha N$ constraints $a$: 
\begin{equation}
    \label{eq:energy-sat}
    E(\vec{\sigma}) =
    \sum\limits_{a=1}^{M} \prod\limits_{i\in \partial a} 
\frac{1-J_a^i \sigma_i}{2}  \ .
\end{equation}
In \eref{eq:energy-sat}, each constraint $a$ affects a set
$\partial a$ of $K$ randomly chosen vertices from the vertex
set $\{1, 2, 3, \ldots, N\}$; $J_a^i$ is
the preferred spin state of constraint $a$ on the vertex
$i \in \partial a$, it takes the quenched value $+1$ or $-1$
with equal probability. If at least one of the vertices
$i \in \partial a$ takes the spin value $\sigma_i = J_a^i$,
the energy of the constraint $a$ is zero, otherwise its energy
is unity. The solution space $\mathcal{S}$ of model
\eref{eq:energy-sat} is formed by all the spin configurations
$\vec{\sigma}$ of zero total energy (i.e., satisfying all the
constraints). In model \eref{eq:energy-sat}, each vertex $i$
is constrained by a set (denoted as $\partial i$) of
constraints.

After the experimental demonstration of a satisfiability
phase-transition in the random $3$-SAT problem by
Kirkpatrick and Selman \cite{Kirkpatrick-Selman-1994},
studies on the solution space structure of the random $K$-SAT
problem have been carried out through rigorous mathematical
methods (see, e.g., 
\cite{Achlioptas-2001,Achlioptas-Naor-Peres-2005,Mezard-etal-2005-a})
and through statistical physics methods (see, e.g.,
\cite{Monasson-Zecchina-1996,Mezard-etal-2002,Mezard-Zecchina-2002,Krzakala-etal-PNAS-2007,Semerjian-2008,Ardelius-Zdeborova-2008}).
The threshold constraint density $\alpha_s(K)$ for the solution space
to be empty is calculated to be $4.2667$ for the random
$3$-SAT problem \cite{Mezard-Zecchina-2002} and its values
for $K\geq 4$ are also predicted by the 1RSB zero-temperature
energetic
cavity method \cite{Mertens-etal-2006}; a lower bound on
$\alpha_s(K)$ is calculated by the zero-temperature
long-range frustration theory \cite{Zhou-2005b,Zhou-2010}.
Ergodicity of the solution space is broken at the clustering
transition point $\alpha = \alpha_d(K) < \alpha_s(K)$
\cite{Mezard-etal-2002,Mezard-Zecchina-2002}. At $\alpha_d(K)
\leq  \alpha \leq \alpha_s(K)$, the solution space
contains an exponential number of Gibbs states. The value of
$\alpha_d(K)$ is calculated by the 1RSB zero-temperature
entropic cavity method in
\cite{Krzakala-etal-PNAS-2007,Montanari-etal-2008},  reporting
$\alpha_d(3)\simeq 3.87$ and $\alpha_d(4) \simeq 9.38$.

We are interested in the heterogeneity of the
ergodic solution space at
$\alpha < \alpha_d(K)$. In the following subsection we calculate
the solution-pair mean overlap as defined by
\eref{eq:overlap} using the replica-symmetric cavity method.
The heterogeneity transition point $\alpha_{cm}(K)$ is determined
by this RS mean-field theory.

\subsection{Replica-symmetric mean-field analysis}

The partition function \eref{eq:Partition_function} is a
weighted sum over all the solution-pairs
$(\vec{\sigma}^1, \vec{\sigma}^2)$. Each vertex is then
associated with a spin vector-state
$(\sigma_i^1, \sigma_i^2)$. Under the coupling field
$x$, we define the cavity probability
$\hat{p}_{a\rightarrow i}(\sigma, \sigma^\prime)$ as the probability that
constraint $a$ is satisfied if the vertex $i \in \partial a$ takes the
spin vector-state $(\sigma, \sigma^\prime)$. Similarly, we define
the cavity probability 
$p_{i\rightarrow a}(\sigma, \sigma^\prime)$ as the probability
that vertex $i$ is in the spin vector-state
$(\sigma, \sigma^\prime)$ in the absence of the constraint
$a \in \partial i$.  

If we exploit the Bethe-Peierls approximation as mentioned in
section~\ref{subsec:kxorsat-rs}, the following RS cavity equations can
be written down for $\hat{p}_{a\rightarrow i}$ and $p_{i \rightarrow a}$
\begin{eqnarray}
\fl \hat{p}_{a\rightarrow i}(\sigma, \sigma^\prime)
	 = 	1- 	\delta_{\sigma}^{-J_a^i}
	 \prod_{j\in \partial a \backslash i}
	\left[\sum_{\sigma_j} p_{j\rightarrow a}(-J_a^j, \sigma_j)\right]
	  -\delta_{\sigma^\prime}^{-J_a^i}
	\prod_{j\in \partial a \backslash i}
	\left[\sum_{\sigma_j} p_{j\rightarrow a}(\sigma_j, -J_a^j)\right]
	\nonumber \\
	   + \delta_{\sigma}^{-J_a^i} \delta_{\sigma^\prime}^{-J_a^i}
	\prod_{j\in \partial a\backslash i}
	 p_{j\rightarrow a}(-J_a^j, -J_a^j) \ ,
 	\label{eq:p_a_to_i} \\
p_{i\rightarrow a}(\sigma, \sigma^\prime)
	=  \frac{e^{x \sigma \sigma^\prime}
	\prod_{b\in \partial i\backslash a}
	\hat{p}_{b\rightarrow i}(\sigma, \sigma^\prime)}{
          \sum\limits_{\sigma_i} \sum\limits_{\sigma_i^\prime}
          e^{x \sigma_i \sigma_i^\prime}
	\prod_{b\in \partial i\backslash a}
	\hat{p}_{b\rightarrow i}(\sigma_i, \sigma_i^\prime)} \ ,
	\label{eq:p_i_to_a}
\end{eqnarray}
where $\delta_m^n$ is the Kronecker symbol ($\delta_m^n=1$ if $m=n$ and
$\delta_m^n=0$ if $m\neq n$).
There is a symmetry requirement for $p_{i\rightarrow a}(\sigma,
\sigma^\prime)$, namely that
$p_{i\rightarrow a}(+1, -1)= p_{i\rightarrow a}(-1,
+1)$. This condition is a result of the fact that, the solution-pair
$(\vec{\sigma}^1, \vec{\sigma}^2)$ has the same contribution to
the partition function \eref{eq:Partition_function} as
the solution-pair $(\vec{\sigma}^2, \vec{\sigma}^1)$.

After a fixed-point has been reached for the RS iterative equations,
the mean value $\overline{q}(x)$
of the solution-pair overlap and the solution-pair 
entropy density $s(q)$ as defined by \eref{eq:entropy} can
then be calculated. For example, the mean solution-pair overlap
is expressed as
\begin{equation}
\label{eq:mean-q-val}
\overline{q}(x)
\equiv \frac{1}{N} \sum\limits_{i=1}^N \langle \sigma_i^1 \sigma_i^2 
\rangle_x
= \frac{1}{N} \sum\limits_{i=1}^{N}
\frac{\sum_{\sigma_i} \sum_{\sigma_i^\prime} 
  \sigma_i \sigma_i^\prime e^{x \sigma_i \sigma_i^\prime}
	\prod_{a\in \partial i}
	\hat{p}_{a\rightarrow i}(\sigma_i, \sigma_i^\prime)}{
          \sum_{\sigma_i} \sum_{\sigma_i^\prime}
          e^{x \sigma_i \sigma_i^\prime}
	\prod_{a\in \partial i}
	\hat{p}_{a\rightarrow i}(\sigma_i, \sigma_i^\prime)} \ ,
\end{equation}
where $\langle \ldots \rangle_x$ means averaging under the
coupling field $x$.

\begin{figure}
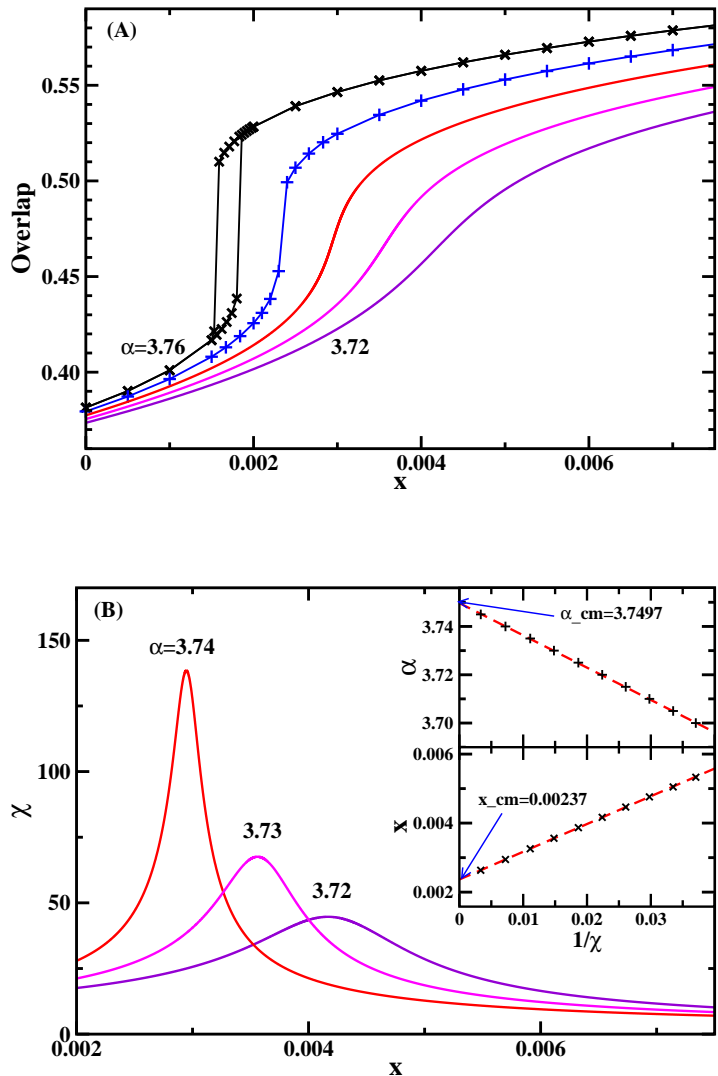

 \begin{center}
	\vskip 0.82cm
        \includegraphics[width=0.6\textwidth]{figure07a.eps}
        \vskip 1.22cm
        \includegraphics[width=0.6\textwidth]{figure07b.eps}
    \end{center}
    \caption{\label{fig:mf-3sat}
     Mean-field predictions on the mean solution-pair overlap
     $\overline{q}(x)$ (A) and the overlap susceptibility $\chi(x)$
     (B) for the random $3$-SAT problem. In (A) the constraint
	density $\alpha$ increases from $3.72$ to $3.76$ (right to
	left) with step	size $0.01$. The inset of (B) shows how
	the inverse of the peak value of $\chi(x)$
	approaches zero as $\alpha$ increases to $\alpha_{cm}(3)=3.7497$
	from below. At $\alpha=\alpha_{cm}$, $\chi(x)$ becomes
	infinite at $x= x_{cm}(3)=0.00237$.
     }
\end{figure}

The predicted mean overlap function $\overline{q}(x)$ for the random
$3$-SAT problem is shown in \fref{fig:mf-3sat}(a) for $\alpha$ in the
vicinity of $3.75$. Similar to the results of the
random $3$-XORSAT problem shown in \fref{fig:q-xorsat-rs}, the
continuity of $\overline{q}(x)$ changes as $\alpha$ exceeds a
critical value $\alpha_{cm}(3) \approx 3.75$. The jumping and 
hysteresis behavior of the mean overlap $\overline{q}(x)$ for $\alpha
> \alpha_{cm}(3)$ indicates that solution communities start to emerge
in the solution space of the random $3$-SAT problem at $\alpha
\approx 3.75$.

The overlap susceptibility $\chi(x)$ measures the sensitivity
of the mean solution-pair overlap with the coupling field $x$,
it is defined by
$\chi(x)  \equiv {\rm d} \overline{q}(x) /{\rm d} x$. The
susceptibility $\chi(x)$ is related to the overlap fluctuation
by
\begin{equation}
    \chi(x) =
    \frac{1}{N}\sum\limits_{i=1}^{N}\sum\limits_{j=1}^{N} \left[
\langle \sigma_i^{1}\sigma_i^{2} \sigma_j^{1} \sigma_j^{2} \rangle_x
    - \langle \sigma_i^{1}\sigma_i^{2} \rangle_x
    \langle \sigma_j^{1} \sigma_j^{2}\rangle_x \right] \ .
	\label{eq:ksat-susceptibility}
\end{equation}
\Fref{fig:mf-3sat}(b) demonstrates that, as $\alpha$ approaches
$3.75$ from below, the peak value of $\chi(x)$ becomes more and
more pronounced and finally diverges.
From the divergence of the peak value
of $\chi(x)$, we obtain
that $\alpha_{cm}(3)=3.7497$ for the random $3$-SAT problem. This
value is much below the value of $\alpha_d(3)=3.87$.

Similar mean-field calculations are performed for the random
$4$-SAT problem and we obtain that $\alpha_{cm}(4)
=8.4746$. This value is again much below the
clustering transition point $\alpha_d(4)=9.38$.

\subsection{Dynamical heterogeneity of Glauber dynamics}

The solution space heterogeneity of the random $K$-SAT problem
influences the dynamics of random walking diffusion processes
\cite{Zhou-Ma-2009}. Similar to the random $K$-XORSAT problem,
we can represent the solution
space of the random $K$-SAT problem as a solution graph of
nodes and edges, with the nodes denoting individual solutions and the 
edges connecting pairs of solutions of unit Hamming distance.
When the constraint density $\alpha$ of the random $K$-SAT problem is less than
$\alpha_d(K)$, this solution graph has a single giant connected component that
includes almost all the solutions of the solution space.
In this ergodic phase of $\alpha < \alpha_d(K)$, the structure of this huge
solution graph become 
heterogeneous when the solutions aggregate into many
different communities. There are many domains of high edge density
in the solution graph, corresponding to the different solution communities.
The nodes of different domains are also connected by many edges, but the
density of inter-domain edges is much lower than the density of intra-domain
edges. As was demonstrated in \cite{Zhou-Ma-2009},
this heterogeneity of edge density causes an entropic trapping
effect to diffusive particles on the solution graph. The dynamics of a
diffusive particle can be decomposed into a trapping
mode (the particle diffuses within a relatively dense-connected
domain of the solution graph) and a transition mode (the particle
escapes from one domain of the solution graph, wonders for a while,
and then enters into another domain of the solution graph). As the trajectory
of the diffusive particle oscillates between the trapping mode and the
transition mode, if a clustering analysis
is performed on a set of solutions sampled from this trajectory at
equal time interval, a clear
community structure can be observed among the sampled solutions
\cite{Zhou-Ma-2009,Zhou-2010}.

The solution space diffusion process can be turned into a stochastic
search algorithm. One such algorithms, the {\tt SEQSAT} of
\cite{Zhou-2009}, constructs a solution for a random
$K$-SAT problem in a sequential manner. Constraints of the problem are
added one after another in a random order, and as a new constraint is
added, a random walk process of single-spin flips is performed in the
solution space of the
satisfied sub-problem to reach a solution that also satisfies the
new constraint. It was observed \cite{Zhou-2009,Zhou-2010} that,
when the constraint density  $\alpha$ of the satisfied sub-problem
exceeds $\alpha_{cm}(K)$, the {\tt SEQSAT} search process
becomes viscous, the mean waiting time needed to
satisfy a new constraints starts to increases rapidly with $\alpha$,
and the sequence of waiting times  starts to have large fluctuations.
This dynamical behavior is easily
understood in terms of the heterogeneity of the underlying solution space.

Solution space structural heterogeneity results in dynamical heterogeneity of
diffusion processes. To demonstrate this point more clearly and to
estimate a
typical relaxation time, we study in this subsection a simple solution space
Glauber dynamics. For a large random $K$-SAT problem with $N$ variables and
$M= \alpha N$ constraints ($\alpha < \alpha_d(K)$), first we construct through
{\tt SEQSAT} a spin configuration that satisfies all the $M$ constraints.
(Of cause we can also use other heuristic algorithms to generate an
initial solution.
The only requirement is that this solution should be a typical solution,
or in other words, it should belong to the single largest solution cluster of
the solution space. This requirement is satisfied in our simulation studies.)
Then we set the initial time as $t=-T_e$ and denote the initial solution as
$\vec{\sigma}(-T_e) \equiv (
\sigma_1(-T_e), \sigma_2(-T_e), \ldots, \sigma_N(-T_e) )$.
The spin configuration
is then updated by single-spin flips at each elementary time step
$\Delta t = 1/N$.
Let us suppose at time $t$ the spin configuration is $\vec{\sigma}(t)$. 
Then a vertex $i$ is chosen uniformly randomly from the whole vertex set
$\{1, 2, \ldots, N\}$; a candidate spin configuration $\vec{\sigma}^\prime$ is
constructed, with $\sigma_j^\prime = \sigma_j(t)$ if $j \neq i$ and
$\sigma_i^\prime = -\sigma_i(t)$. If $\vec{\sigma}^\prime$ is not a
solution
of the $K$-SAT problem, then at time $t^\prime=t + \Delta t$, the old
spin configuration is
kept, i.e., $\vec{\sigma}(t^\prime)=\vec{\sigma}(t)$. However, if
$\vec{\sigma}^\prime$ is also a solution, then with probability
one-half $\vec{\sigma}(t^\prime)
= \vec{\sigma}(t)$ and with the remaining probability one-half
$\vec{\sigma}(t^\prime) = \vec{\sigma}^\prime$.  

\begin{figure}
 \begin{center}
	\vskip 0.22cm
        \includegraphics[width=0.6\textwidth]{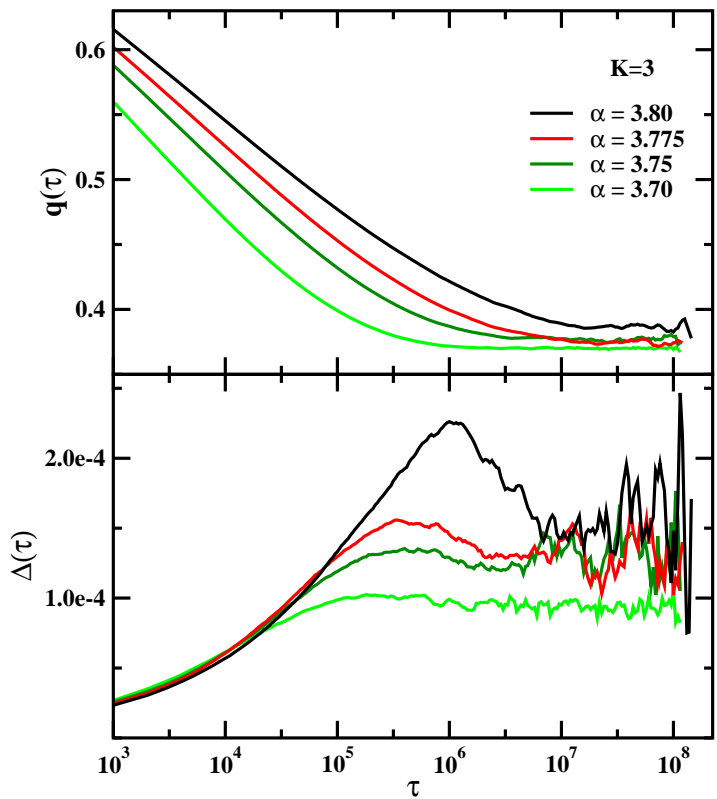}
    \end{center}
    \caption{\label{fig:overlap-3sat}
     Glauber dynamics simulation results for a single random $K$-SAT ($K=3$)
     problem instance with $N=10^5$ vertices. At a given value of $\alpha$,
     the first $M= \alpha N$ constraints of this single random instance
      are used in the simulation.
     (Upper panel) Mean value of the overlap $q(\tau)$
     between two solutions separated by time $\tau$ along the trajectory
     of configuration evolution. 
     (Lower panel) The variance $\Delta(\tau)$ of the solution-pair
	overlap $q(\tau)$.
     }
\end{figure}

\begin{figure}
 \begin{center}
        \includegraphics[width=0.6\textwidth]{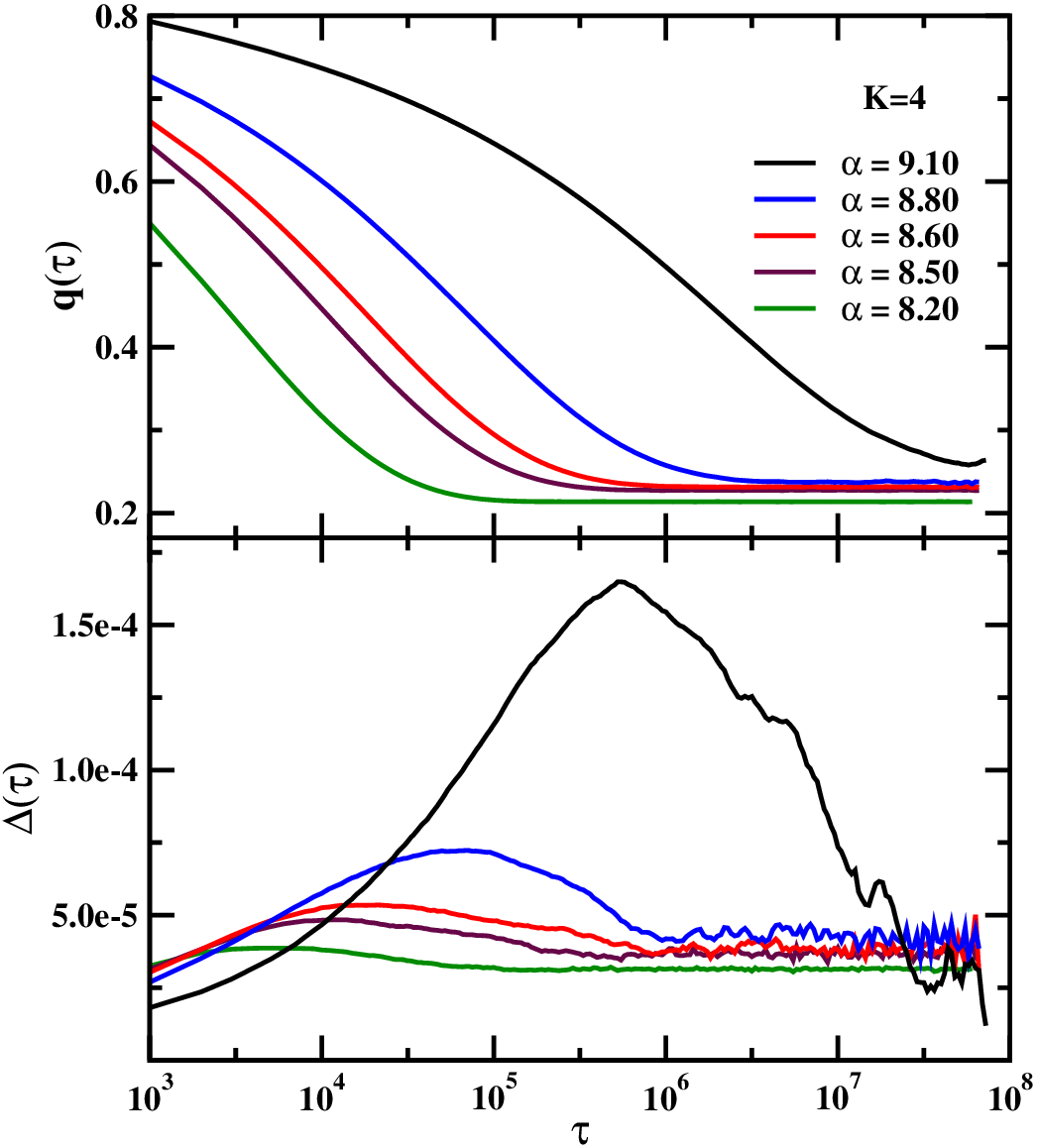}
    \end{center}
    \caption{\label{fig:overlap-4sat}
     Same as \fref{fig:overlap-3sat}, but for $K=4$.
     }
\end{figure}

A unit time of the above-mentioned Glauber dynamics corresponds to
$N$ spin-flip attempts.
The actual number of accepted spin flips in a unit time is about
$0.1 N$ for the problem
instances of \fref{fig:overlap-3sat} and \fref{fig:overlap-4sat}.
If this random walk diffusion process is simulated
for an extremely long period of time, every solution in the
connected component of the
solution graph which the initial solution $\vec{\sigma}(-T_e)$ belongs
will have the same frequency of being visited. The time $T_e$ is
set to be large enough (e.g., $T_e \sim 10^8$) to ensure that
the diffusion process has completely forget the initial
solution $\vec{\sigma}(-T_e)$ at time $t\geq 0$.

For the trajectory of spin configurations at $t\geq 0$, the quantity
$q(\tau) =(1/N)\sum_{i=1}^{N} \sigma_i(t) \sigma_i(t+\tau)$
measures the overlap between two spin configurations $\vec{\sigma}(t)$ and
$\vec{\sigma}(t+\tau)$ that are separated by a time $\tau$. $q(\tau)$
is a random variable, its value fluctuates with
different choices of the time $t$. 
We are mainly interested in the mean value $\overline{q}(\tau)$ and
the variance $\Delta(\tau)$ of the overlap $q(\tau)$.
The mean overlap $\overline{q}(\tau)$ is calculated by
\begin{equation}
 \overline{q}(\tau) \equiv  \langle q(\tau)\rangle_t
 =\frac{1}{N} \sum\limits_{i=1}^{N} \langle \sigma_i(t) \sigma_i(t+\tau) 
 \rangle_t \ ,
\label{eq:ksat-mo}
\end{equation}
where $\langle \ldots \rangle_t$ means averaging over different
starting times $t\geq 0$
along the diffusion trajectory.
The variance $\Delta(\tau)$ is expressed as
\begin{eqnarray}
\fl \Delta(\tau) = \frac{1}{N^2} 
\sum_{i=1}^{N} \sum_{j=1}^{N}
\left[ \langle \sigma_i(t) \sigma_i(t+\tau)  \sigma_j(t)  \sigma_j(t+\tau)
\rangle_t - \langle \sigma_i(t) \sigma_i(t+\tau) \rangle_t
 \langle \sigma_j(t) \sigma_j(t+\tau) \rangle_t
\right] \ .  \nonumber \\
\; \label{eq:ksat-fluctuation}
\end{eqnarray}
Comparing \eref{eq:ksat-fluctuation} with \eref{eq:ksat-susceptibility},
a dynamical susceptibility $\chi_4(\tau)$ can be defined as
$\chi_4(\tau) \equiv N \Delta(\tau)$ 
(such a quantity was introduced in \cite{Parisi-1997,Donati-etal-2002},
see also the review \cite{Cavagna-2009}).

The upper panel of \fref{fig:overlap-3sat} shows the relaxation behavior
of $\overline{q}(\tau)$ for a large random $3$-SAT formula with $N=10^5$
vertices. To study how the shape of $\overline{q}(\tau)$ changes with 
constraint density $\alpha$, the first $M= \alpha N$ constraints of this
same formula is used for Glauber dynamics simulation at each value
of $\alpha$. We focus on $\alpha$ values in the vicinity of $\alpha_{cm}(3)$.
For $\alpha=3.70$, the mean overlap $\overline{q}(\tau)$
reaches its plateau value of $0.37$ at $\tau \approx 10^6$.
For $\alpha=3.75 \simeq \alpha_{cm}(3)$, the plateau of
$\overline{q}(\tau)$ is reached at $\tau \approx 10^{6.5}$, and its value
increases to $0.38$. For $\alpha=3.80$, the plateau of
$\overline{q}(\tau)$ is reached at $\tau \approx 10^{7.2}$, with a value of
about $0.39$.  These and our other unshown simulation
results clearly confirm that, as $\alpha$ exceeds
$\alpha_{cm}(3)$, the relaxation of the solution space diffusion process
is slowed down greatly. 

The lower panel of \fref{fig:overlap-3sat} shows the variance $\Delta(\tau)$
of the solution-pair overlap values.
At $\alpha=3.70$, the variance $\Delta(\tau)$ increases with $\tau$ and
reaches its plateau value at $\tau \approx 10^6$. At $\alpha=3.75$,
$\Delta(\tau)$ starts to show a peak at $\tau \approx
10^{5.5}$. The peak of $\Delta(\tau)$ becomes more and more pronounced as
$\alpha$ further increases, and the time $\tau_{dh}$, corresponding to
the peak value of $\Delta(\tau)$, shifts to larger values with $\alpha$. At
$\alpha=3.80$, $\tau_{dh}\approx 10^{6}$.

The peak time $\tau_{dh}$ of the variance $\Delta(\tau)$ gives a 
measure of the typical time scale of dynamical heterogeneity of the Glauber
dynamics. For $\tau \ll \tau_{dh}$, two solutions $\vec{\sigma}(t)$ and
$\vec{\sigma}(t+\tau)$ have a large chance of being in the 
same solution community,
therefore the fluctuation of overlap values $q(\tau)$ is small; a
large fraction of vertices are inactive in this relatively short
time window of $\tau$, as their spin values are flipped only with low
frequencies. On the other hand, for $\tau \gg \tau_{dh}$, the compared two
solutions $\vec{\sigma}(t)$ and $\vec{\sigma}(t+\tau)$ have a large chance
of belonging to two different solution communities, and therefore the
fluctuation of their overlap values is again small; the spin values of
most of the vertices have been flipped many times during
such a large time window, and hence dynamical heterogeneity is destroyed.
When the time window $\tau$ is comparable to $\tau_{dh}$, 
the solutions $\vec{\sigma}(t)$ and $\vec{\sigma}(t+\tau)$ have comparable
probabilities of being in the same solution community and being in two
different communities; this causes relatively large fluctuation of the
overlap values. As $\alpha$ increases from $\alpha_{cm}(3)$ further, it
takes more time for the diffusion process to escape from a solution community,
and the
difference between intra- and inter-community overlap values become larger,
these two facts make the  peak value $\Delta(\tau)$ and the peak time
$\tau_{dh}$ both to increase rapidly.

\Fref{fig:overlap-4sat} reports the simulation results for a random $4$-SAT
problem instance of $N=10^5$ vertices and $M = \alpha N$ constraints.
Similar to the case of random $3$-SAT, we find that the viscosity of the
Glauber dynamics increases rapidly with constraint density $\alpha$. For
$\alpha=9.10> \alpha_{cm}(4)$, the mean overlap
$q(\tau)$ reaches its plateau value only at a very large value of
$\tau \approx 10^8$. As demonstrated by the variance $\Delta(\tau)$
of solution-pair overlaps, dynamical heterogeneity become to manifest itself
at $\alpha = 8.50 \simeq \alpha_{cm}(4)$ and it becomes more and more
pronounced as $\alpha$ further increases.

Comparing \fref{fig:overlap-3sat} and \fref{fig:overlap-4sat} we can
also notice an important dynamical difference between the random 
$3$-SAT and the random $4$-SAT case.
For the random $4$-SAT system, we observe that the relaxation curve
of $\overline{q}(\tau)$ is $`$Z' form in shape at
$\alpha=9.10$ (i.e., $97\%$ of $\alpha_d(4)$), with a plateau value
of $q\approx 0.7$ at intermediate time intervals $\tau$ before it
finally decay to a much lower $q\approx 0.3$ at $\tau \sim 10^8$.
The peak of the overlap variance $\Delta(\tau)$ corresponds
to the time interval $\tau$ at which $\overline{q}$ starts to decay
from the larger plateau. 
However, for the random $3$-SAT system, we find that the
curve of $\overline{q}(\tau)$ has a $`$L' shape even at
$\alpha=3.80$ ($98\%$ of $\alpha_d(3)$), and at the time when
$\Delta(\tau)$ reaches its peak value, the mean overlap
$\overline{q}(\tau)$ is already close to its asymptotic value of
$\tau \rightarrow \infty$. We suggest that this dynamical difference
is a strong reflect of the difference between the community structures of
the random $3$-SAT and the general random $K$-SAT problems with $K\geq 4$.

For the random $3$-SAT problem, in the heterogeneity phase of
$\alpha_{cm}(3) < \alpha<\alpha_d(3)$, the solution space probably is
dominated by a sub-exponential number of largest solution communities.
Because of this predominance, the mean intra-community solution-pair
overlap of each of these dominating communities is only slightly higher
than the mean solution-pair overlap of the whole solution space. Then
at typical time $\tau_{dh}$ when the diffusive particle jumps between
different dominating solution communities, one will not observe much
drop in the mean overlap value $\overline{q}$. Such a picture is
consistent with the prediction that, at $\alpha> \alpha_{d}(3)$,
only a sub-exponential number of solution Gibbs states
dominate the solution space \cite{Krzakala-etal-PNAS-2007}.

For the random $K$-SAT problem with $K\geq 4$, however,
in the range of $\alpha \in (\alpha_{cm}(K), \alpha_d(K))$ the
solution space probably is contributed mainly by a exponential
number of median-sized solution communities. The sub-exponential number
of the largest solution communities has only a negligible contribution
to the whole solution space and therefore barely influence the
dynamics of the diffusive particle. For such a community structure,
then the mean intra-community solution-pair overlap will be much
larger than the mean solution-pair overlap of the whole solution
space, resulting in a change of trend of $\overline{q}(\tau)$ at
$\tau \sim \tau_{dh}$. Such a picture for $K\geq 4$ is again
consistent with the prediction that, at $\alpha> \alpha_d(K)$ the
solution space is dominated by an exponential number of median-sized
Gibbs states \cite{Krzakala-etal-PNAS-2007}.

\subsection{Summary for $K$-SAT}

In this section, we confirmed that the
solution space of the random $K$-SAT problem becomes heterogeneous
at $\alpha \geq \alpha_{cm}(K)$ and determined by the
replica-symmetric cavity method that $\alpha_{cm}(3)\simeq 3.75$ and 
$\alpha_{cm}(4)\simeq 8.47$. We demonstrated that the existence of many
solution communities in the solution space caused 
heterogeneous behavior in the dynamics of a solution space
diffusion process. The typical time scale of dynamical
heterogeneity of this diffusion process is determined
by computer simulations.

\section{Outlook}
\label{sec:discuss}

A heterogeneity transition was found to occur in the
ground-state configuration spaces of 
two multiple-spin interaction systems, the random $K$-XORSAT problem 
and the random $K$-SAT problem. We expect that this
transition is a general phenomenon
that occurs in many other spin glass systems before the
ergodicity-breaking transition of the ground-state configuration
space. Such a heterogeneity
transition is unlikely to be special to the ground-state
configuration space but should also be observed as the energy level
(or equivalently the temperature $T$) of the configuration space
is lowered to certain critical level. If the configuration space of a
spin glass system is ergodic but highly heterogeneous at certain
temperature $T$, this structural heterogeneity probably will
manifest itself through heterogeneous behaviors in various
spin relaxation dynamical processes of the system. More deep
understanding on the relationship between the phenomenon of
dynamical heterogeneity and the structural heterogeneity of
the configuration space is very desirable. Such efforts may bring
new ways of probing configuration space heterogeneity from
observing features of dynamical heterogeneity.

The solution space heterogeneity of the random $K$-SAT problem has
been studied analytically only through the replica-symmetric cavity
method. However, from the experiences gained on the random $K$-XORSAT
problem, we believe the replica-symmetric cavity
theory is not sufficient for a heterogeneous solution space.
A complete study on the heterogeneity transition of the random $K$-SAT
using the 1RSB mean-field cavity theory will be reported in a later
paper.

In the case of the random $K$-SAT problem, we have not yet
performed a systematic investigation on the
scaling behaviors of the peak value of overlap variance
(equivalently, the overlap susceptibility $\chi(\tau)$)
and the typical time $\tau_{dh}$ of dynamical heterogeneity.
Probably both the peak value of $\chi(\tau)$ and the characteristic
time $\tau_{dh}$ diverge at the clustering transition point
$\alpha=\alpha_d(K)$.
To get unambiguous results, we need a more
efficient protocol of simulating the diffusion dynamics on
the solution space.

Simulations results of \cite{Zhou-Ma-2009} and \cite{Li-Ma-Zhou-2009}
indicated that, for the random $3$-SAT and random $4$-SAT problem,
the single solution space Gibbs states at $\alpha> \alpha_d(K)$
are themselves very heterogeneous in internal structure. To study
analytically the heterogeneity of single solution clusters in the
ergodicity-breaking phase, however, appears to be a very
challenging task. Such kind of investigations may be very valuable
for us to understand how the largest solution clusters of the
random $K$-SAT problem evolve with constraint density $\alpha$.

\section*{Acknowledgement}

HZ thanks Dr. Lenka Zdeborova for helpful correspondences;
CW thanks Professor Taiyu Zheng for her kind support.
This work was partially supported by the
National Science Foundation of China
(Grant numbers 10774150 and 10834014) and the
China 973-Program (Grant number 2007CB935903).

\appendix

\section{The 1RSB cavity equations for the random $K$-XORSAT problem}
\label{sec:appendix-a}

Here we list the technical details of the 1RSB mean-field cavity
theory  for the random $K$-XORSAT problem as discussed in
section \ref{sec:1rsb-xorsat}. For a comprehensive description of the
1RSB mean-field theory, the reader is referred to
\cite{Mezard-Parisi-2001,Mezard-Montanari-2006,Krzakala-etal-PNAS-2007,Montanari-etal-2008}.

When there are an exponential number of Gibbs states, the probability
$p_{i\rightarrow a}^+$ as defined in section
\ref{subsec:kxorsat-rs} will be different
for different Gibbs states. The distribution of
$p_{i\rightarrow a}^+$ among all the Gibbs states is denoted by
$P_{i\rightarrow a} ( p_{i\rightarrow a}^+)$.
We define $\overline{p}_{i\rightarrow a}^+$ as
the average value of $p_{i\rightarrow a}^+$, 
i.e.,  $\overline{p}_{i\rightarrow a}^+ = \int {\rm d} p^+
P_{i\rightarrow a}(p^+) p^+$. We also define two auxiliary
distributions
as \cite{Mezard-Montanari-2006}
\begin{eqnarray}
 Q_{i\rightarrow a}^{+}(p_{i\rightarrow a}^+ | 
\overline{p}_{i\rightarrow a}^+)
 & = \frac{ P_{i\rightarrow a}( p_{i\rightarrow a}^+ ) p_{i\rightarrow a}^+ }{
\overline{p}_{i\rightarrow a}^+} \ , \\
 Q_{i\rightarrow a}^{-}(p_{i\rightarrow a}^+ | \overline{p}_{i\rightarrow a}^+)
& = \frac{ P_{i\rightarrow a}( p_{i\rightarrow a}^+ ) (1-p_{i\rightarrow a}^+) }{
(1- \overline{p}_{i\rightarrow a}^+)} \ .
\end{eqnarray}
$Q_{i\rightarrow a}^+ ( p_{i\rightarrow a} |
\overline{p}_{i\rightarrow a}^+)$ is the conditional probability of
$p_{i\rightarrow a}^+$ given that $\sigma_i = +1$.
Similarly,  $Q_{i\rightarrow a}^- ( p_{i\rightarrow a} |
\overline{p}_{i\rightarrow a}^+)$ is the conditional probability of
$p_{i\rightarrow a}^+$ given that $\sigma_i = -1$.

At the special case of $m=1$, it can be shown that the iteration equation for
$\overline{p}_{i\rightarrow a}^+$ has the same form as
\eref{eq:bp-xorsat}, with the only difference that all the
$p_{j\rightarrow b}^+$ values are replaced by their
corresponding mean $\overline{p}_{j\rightarrow b}^+$, i.e.,
$\overline{p}_{i\rightarrow a}^+ = \hat{p}(\{
\overline{p}_{j\rightarrow b}^+ \})$ \cite{Krzakala-etal-PNAS-2007}.
The iteration equations for the conditional
probabilities $Q_{i\rightarrow a}^{+}(p_{i\rightarrow a}^+ |
 \overline{p}_{i\rightarrow a}^+)$
 and
 $Q_{i\rightarrow a}^{-}(p_{i\rightarrow a}^+ |
 \overline{p}_{i\rightarrow a}^+)$
at $m=1$ have the following expression
\cite{Mezard-Montanari-2006,Krzakala-etal-PNAS-2007}
\begin{eqnarray}
\fl Q_{i\rightarrow a}^{\sigma_i}(p_{i\rightarrow a}^+ | 
\overline{p}_{i\rightarrow a}^+)
= \prod\limits_{b\in \partial i \backslash a}
\left[
\sum\limits_{\sigma_{\partial b \backslash i}}
w_{b\rightarrow i}^{\sigma_i} ( \sigma_{\partial b \backslash i} )
\prod\limits_{j\in \partial b\backslash i}
Q_{j\rightarrow b}^{\sigma_j} (p_{j\rightarrow b}^+
 | \overline{p}_{j\rightarrow b}^+) 
\right]
\delta ( p_{i\rightarrow a}^+ -
\hat{p}( \{p_{j\rightarrow b}^+ \}) ) \ ,
\nonumber \\
 \ 
\end{eqnarray}
where $\sigma_{\partial b\backslash i}$ denotes a spin configuration for
the vertex set $\partial b\backslash i$ of constraint $b$, and 
\begin{equation}
w_{b\rightarrow i}^{\sigma_i}
(\sigma_{\partial b \backslash i} ) = 
\frac{ \prod\limits_{j\in \partial b\backslash i}
 \overline{p}_{j\rightarrow b}^{\sigma_j}
 \delta_{\sigma_i \prod_{j \in
b \backslash i} \sigma_j}^1 }{
\sum\limits_{\sigma_{\partial b\backslash i}}
\prod\limits_{j\in \partial b\backslash i}
 \overline{p}_{j\rightarrow b}^{\sigma_j} \delta_{\sigma_i \prod_{j \in
b \backslash i} \sigma_j}^1 }
\end{equation}
is the probability of a satisfying spin assignment
$\sigma_{\partial   b\backslash i}$ for constraint $b$
given the spin value $\sigma_i$
of vertex $i$ (the probability $\overline{p}_{j\rightarrow b}^{-}
\equiv 1- \overline{p}_{j\rightarrow b}^+$ is
the mean probability of vertex $j$ taking spin value $\sigma_j=-1$
in the absence of constraint $b$).

The cavity iterative equations for $\overline{p}_{i\rightarrow a}^+$
and $Q_{i \rightarrow a}^{\sigma_i}( p_{i \rightarrow a}^+ |
\overline{p}_{i\rightarrow a}^+)$
can be solved by the population dynamics technique
\cite{Mezard-Parisi-2001,Mora-Mezard-2006}. We are interested in
the solution space property at a fixed value of the overlap value
$q$, so in the population dynamics simulation the magnitude of
the coupling field $x$ is adjusted from time to time to ensure
that the mean overlap value as expressed by \eref{eq:q-xorsat-rs} with
$p_{j\rightarrow a}^+$ replaced by $\overline{p}_{j\rightarrow a}^+$
is equal to the specified value $q$ (see, e.g., 
\cite{Krzakala-RicciTersenghi-Zdeborova-2009,Mora-Mezard-2006}).

The grand free-energy density $g$ of the model system, as defined by
$g=(1/N) \ln Z_{1RSB}$, has the following simplified expression at
$m=1$:
\begin{eqnarray}
  g = & 
\frac{1}{N} \sum\limits_{i=1}^{N}
\ln\left( \sum\limits_{\sigma_i} e^{x \sigma_i}
 \prod\limits_{a\in \partial i}
\frac{1+ \sigma_i \prod_{j\in \partial a\backslash i} 
(2 \overline{p}_{i\rightarrow a}^+ -1)}{2} \right) \nonumber \\
& - \frac{1}{N} \sum\limits_{a=1}^{\alpha N}(K-1)
\ln\left(\frac{1+\prod_{i\in \partial a} 
(2\overline{p}_{i\rightarrow
    a}^+ -1)}{2}
\right)
\end{eqnarray}
The mean free energy density $\overline{f}$ of a Gibbs state is
expressed as
\begin{equation}
  \overline{f} = \frac{1}{N}\sum\limits_{i=1}^{N} \overline{\Delta
    f_i} - \frac{1}{N} \sum\limits_{a=1}^{\alpha N} (K-1)\overline{\Delta
    f_a} \ ,
\end{equation}
where $\overline{\Delta f_i}$ and $\overline{\Delta f_a}$ are,
respectively, the free energy increase caused by vertex $i$ and
constraint $a$. These two free energy increases are expressed by the
following expressions at $m=1$:
\begin{eqnarray}
  \overline{\Delta f_i}
  = & \sum\limits_{\sigma_i} \overline{p}_i^{\sigma_i}
\prod\limits_{a\in \partial i} \left[
\sum\limits_{\sigma_{\partial a \backslash i}} w_{a\rightarrow
  i}^{\sigma_i}(\sigma_{\partial a\backslash i}) 
\prod\limits_{j\in \partial a\backslash i} Q_{j\rightarrow
  a}^{\sigma_j}(p_{j\rightarrow a}^+ | \overline{p}_{j\rightarrow
  a}^+) \right] \times \nonumber \\
&  \ln\left(
\sum\limits_{\sigma_i} e^{x \sigma_i} \prod\limits_{a\in \partial i}
\frac{1+ \sigma_i \prod_{j\in \partial a\backslash i} (2
  p_{j\rightarrow a}^+ -1)}{2} \right)  \ ,
\label{eq:d-f-i}
\end{eqnarray}
and
\begin{equation}
\fl  \overline{\Delta f_a}
  = \sum\limits_{\sigma_{\partial a}} \left[ w_a(\sigma_{\partial a})
\prod\limits_{i\in \partial a} Q_{i\rightarrow a}^{\sigma_i}(
p_{i\rightarrow a}^+ | \overline{p}_{i\rightarrow a}^+) \right]
\ln\left(\frac{1+\prod_{i\in \partial a} (2 p_{i\rightarrow a}^+
  -1)}{2}\right)  \ .
	\label{eq:d-f-a}
\end{equation}
In \eref{eq:d-f-i}), $\overline{p}_i^{+}$ is expressed as
\begin{equation}
\overline{p}_{i}^{+}=
 = 
 \frac{ e^{x} \prod\limits_{a\in \partial i}
\left[ \frac{1+ \prod\limits_{j\in \partial a \backslash i}
 (2 \overline{p}_{j\rightarrow a}^+ -1)}{2} \right]}{
 e^{x} \prod\limits_{a\in \partial i} \left[
\frac{1+ \prod\limits_{j\in \partial a \backslash i}
 (2 \overline{p}_{j\rightarrow a}^+-1)}{2} \right]
+  e^{-x} \prod\limits_{a\in \partial i} \left[
\frac{1- \prod\limits_{j\in \partial a\backslash i}
 (2 \overline{p}_{j\rightarrow a}^+ -1)}{2} \right]
} \ ,
\end{equation}
and $\overline{p}_i^- = 1- \overline{p}_i^+$. The 
probability $w(\sigma_{\partial a})$ in \eref{eq:d-f-a} is expressed
as
\begin{equation}
w_{a}(\sigma_{\partial a} ) = 
\frac{ \prod\limits_{i \in \partial a}
 \overline{p}_{i\rightarrow a}^{\sigma_i}
 \delta_{ \prod_{i \in
a} \sigma_i}^1 }{
\sum\limits_{\sigma_{\partial a}}
\prod\limits_{i\in \partial a}
 \overline{p}_{i\rightarrow a}^{\sigma_i} \delta_{\prod_{i \in
a} \sigma_i}^1 } \ .
\end{equation}

At $m=1$, the complexity $\Sigma$, the total free entropy density
$s_{total}$, and the mean entropy density of a solution community,
$s_{cm}$, are calculated to be
\begin{eqnarray}
  \Sigma = & g - \overline{f} \ , \\
  s_{total} = & g - x q \ , \\
  s_{cm} = & \overline{f} - x q \ .
\end{eqnarray}
The equality $s_{total}= s_{cm} + \Sigma$ holds at $m=1$.

It appears that at some overlap values $q$, more than one mean-field
1RSB solutions can be produced by the population dynamics simulation
 at $m=1$. These different mean-field solutions have the same value of
$s_{total}$ but different values of $\Sigma$ and $s_{cm}$. In such a
case we choose the mean-field solution with the largest value of
$\Sigma$ (similar situations of multiple mean-field solutions were
observed earlier in the random $K$-SAT problem at $m<1$
\cite{Zhou-2008,Montanari-etal-2008}).

\noappendix

\section*{References}


\end{document}